\documentclass[preprint,review,12pt]{elsarticle}

\newtheorem{proposition}{Proposition}

\newtheorem{example}{Example}

\newcommand{\cv}[1]{\overline{#1}\,}

\newcommand{\BB}{\mathbb B}

\newcommand{\BDT}{{\mathbb T}}

\newcommand{\bone}{{\mathbf 1}}
\newcommand{\bzero}{{\mathbf 0}}

\newcommand{\bx}{{\mathbf x}}
\newcommand{\by}{{\mathbf y}}
\newcommand{\ba}{{\mathbf a}}

\newcommand{\bc}{{\mathbf c}}

\newcommand{\bp}{{\mathbf p}}

\newcommand{\be}{{\mathbf e}}

\newcommand{\cS}{{\mathbf S}}

\newcommand{\cI}{{\mathbf I}}
\newcommand{\cJ}{{\mathbf J}}
\newcommand{\cT}{{\mathbf T}}

\newcommand{\ga}{\alpha}

\usepackage{comment}
\usepackage{tikz}
\usepackage[utf8]{inputenc}
\usepackage[T1]{fontenc}
\usetikzlibrary{positioning}
\usetikzlibrary{positioning, shapes.geometric}
\usepackage{caption}
\usepackage{array}
\usepackage{geometry}
\usepackage{float}

\journal{Discrete Applied Mathematics}

\begin{document}

\begin{frontmatter}



\title{Sequential testing problem: A follow-up review}


\author{Tonguç Ünlüyurt}
 \ead{tonguc@sabanciuniv.edu}
\affiliation{organization={Sabanci University},
            addressline={Orta mah. Üniversite cad. No:27}, 
            city={Istanbul},
            postcode={34956}, 
            country={Türkiye}}

\begin{abstract}
This review aims to provide a comprehensive update on the progress made on the Sequential Testing problem (STP) in the last 20 years after the review, \cite{unluyurt2004sequential} was published. Many studies have provided new theoretical results,  extensions of the problem, and new applications. In this review, we pinpoint the main results and discuss the relations between the problems studied. We also provide possible research directions for the problem.
\end{abstract}



\begin{keyword}
Sequential testing problem \sep stochastic function evaluation \sep search problem \sep network connectivity \sep minimum cost diagnosis


\end{keyword}

\end{frontmatter}



\section{Introduction}
\label{intro}

This review is a follow up for \cite{unluyurt2004sequential} where  a review is provided for the STP. There have been numerous studies since then studying different models, solution approaches, and applications. In this article, we put together and review all these studies published after then. We also discuss some references that were missing in the previous review.

In the STP, we try to find a best strategy to evaluate a given Boolean (or a general discrete) function. The values of the variables are unknown to start with, so we need to learn the values of some of the variables that will allow us to determine the correct value of the function. Learning the value of a variable has an associated cost.  On the other hand, we have some probabilistic information regarding the values of the variables. So, a feasible solution for the problem is a strategy that outputs the next variable to learn or the correct value of the function. Since there are costs involved, the ultimate goal is to develop a best strategy that optimizes a certain cost metric (typically the total expected cost).

The problem has been studied for a wide range of applications, including network connectivity, database querying, search problems, diagnosis problems, and others. New extensions and variations of the problem, sometimes with common characteristics, have been introduced and studied.

The remainder of the paper is organized as follows. We will describe the problem setting and recall certain classes of Boolean functions in section \ref{probdef} along with some important issues regarding the representation of the problem instances. Then we will review the relevant literature in terms of problem setting and/or the application in section \ref{rev}. We will conclude by providing a short summary and identifying some future research directions in Section \ref{conclusion}.

\section{Problem Setting} \label{probdef}

In this section, we will provide a general setting of the problem. Since this is a follow-up review, the problem description will be rather short compared to \cite{unluyurt2004sequential} and we refer the reader to \cite{unluyurt2004sequential} for a more detailed discussion. We use slightly different terminology than \cite{unluyurt2004sequential} which is quite straightforward to match. 

We are given a Boolean function $f:\{0,1\}^n\mapsto \{0,1\}$ that describes the functionality of a system, the status of a patient, the result of a query, state of the system, etc.  The goal is to evaluate the function for the current values of the variables $\textbf{x}=(x_1,x_2,...,x_n)$ that are unknown. It is possible to learn the values of the variables by paying the associated cost. In particular, it costs $c_i>0$ to learn the value of $x_i$. It may be the case that this cost is the cost of a medical test, a mechanical test, the time it takes to learn the value of the variable over a distributed database, etc. In addition, we have probabilistic information regarding the value of each variable. In particular, the probability that the variable $x_i$ takes the value 1 is $p_i$ where $p_i+q_i=1$. In many applications and studies, it is assumed that the variables take values independent of each other.  

In order to learn the value of the function at the current state, we need to learn the values of some of the variables that will enable us to conclude about the correct value of the function. The values of the variables are learned one by one and in general the next variable to learn may depend on the values of previously learned variables. We will refer to such a solution as a \emph{strategy}. So a \emph{strategy} either determines the next variable to learn or outputs the correct value of the function given the values of the variables whose values are learned so far.  

We can associate an expected cost for each \emph{strategy} since the cost of learning the variables can be computed for any realization of values for the variables. In addition, we can compute the probability of any realization of values for the variables. One can compute the expected cost of a \emph{strategy} in different ways, sometimes exploiting the special structure of the function or the \emph{strategy} itself. One general approach is to consider all possible vectors $\textbf{x} \in \{0,1\}^n$. The probability of realization is $\textbf{x}$ is $p(\textbf{x}) = \prod_{i:x_i=1}p_i \prod_{i:x_i=0}q_i$. Letting $c_S(\textbf{x})$ as the cost incurred by the strategy $S$, one can calculate the expected cost of a strategy $EC(S)$ as follows:
$$ EC(S)=\sum_{\textbf{x} \in \{0,1\}^n} p(\textbf{x})c_S(\textbf{x}).$$

The STP involves finding a \emph{strategy} with the minimum expected cost. Since learning the value of a variable has a positive cost, in a \emph{optimal strategy} we should stop as soon as we can determine the value of the function by fixing the values of the variables that have been learned so far. 

A \emph{strategy} can naturally be represented as a Binary Decision Tree (BDT). The internal nodes of the tree are labeled with the variables, and each node other than the leaves has two children corresponding to the associated variable being 1 or 0.The first variable to learn is the variable at the root node. Depending on the value of the variable, one can continue learning the value of the next variable or conclude that the function is 0 or 1, if we reach a leaf node. The leaf nodes are labeled 0 or 1 indicating the correct value of the function. In fact, the BDT can be interpreted as classifying the incoming binary vectors into two classes. 

\begin{example}\label{cr1e1.1}
We consider the following function with three variables.

\begin{equation}\label{cr11}
f(x_1,x_2,x_3)=(x_1 \wedge x_2)\vee (x_2 \wedge x_3)
\end{equation}
Then, Figure \ref{ins} shows a
possible strategy, in which the value of  $x_1$ is
learned first, and depending on whether it is 1 or 0,
variables $x_2$ or $x_3$ are inspected next, etc.  For example, if the
system is in state $\bx=(1,1,0)$, then this strategy
learns that the value of the function is 1 after learning the values of
$x_1$ and $x_2$, while if the system's state is $\bx=(0,1,0)$,
then this strategy will stop after learning the values of variables $x_1$ and $x_3$, concluding that the value of the function is 0.
\end{example}
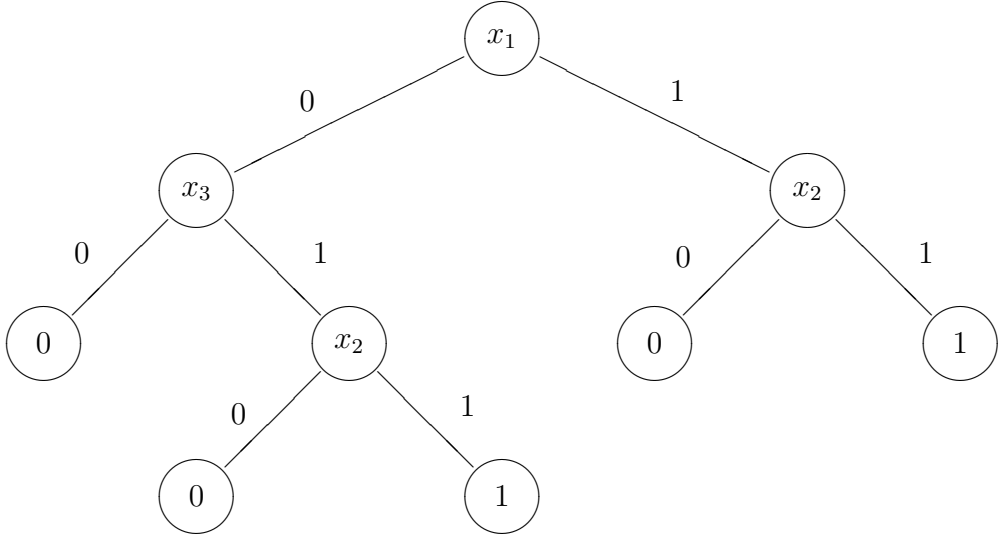
\begin{figure}[htbp]
\begin{center}
\setlength{\unitlength}{0.01in}
\begin{picture}(520,280)(100,560)
\thinlines \put(520,740){\circle{40}} \put(200,740){\circle{40}}
\put(280,660){\circle{40}} \put(120,660){\circle{40}}
\put(440,660){\circle{40}} \put(600,660){\circle{40}}
\put(360,580){\circle{40}} \put(200,580){\circle{40}}
\put(360,820){\circle{40}} \put(340,810){\line(-2,-1){120}}
\put(380,810){\line( 2,-1){120}} \put(535,725){\line( 1,-1){ 50}}
\put(505,725){\line(-1,-1){ 50}} \put(215,725){\line( 1,-1){ 50}}
\put(295,645){\line( 1,-1){ 50}} \put(185,725){\line(-1,-1){ 50}}
\put(265,645){\line(-1,-1){ 50}}
\put(360,815){\makebox(0,0)[b]{$x_1$}}
\put(200,735){\makebox(0,0)[b]{$x_3$}}
\put(520,735){\makebox(0,0)[b]{$x_2$}}
\put(600,655){\makebox(0,0)[b]{$1$}}
\put(440,655){\makebox(0,0)[b]{$0$}}
\put(280,655){\makebox(0,0)[b]{$x_2$}}
\put(360,575){\makebox(0,0)[b]{$1$}}
\put(200,575){\makebox(0,0)[b]{$0$}}
\put(120,655){\makebox(0,0)[b]{$0$}}
\put(452,787){\makebox(0,0)[b]{$1$}}
\put(258,782){\makebox(0,0)[b]{$0$}}
\put(265,702){\makebox(0,0)[b]{$1$}}
\put(140,702){\makebox(0,0)[b]{$0$}}
\put(582,702){\makebox(0,0)[b]{$1$}}
\put(455,700){\makebox(0,0)[b]{$0$}}
\put(342,622){\makebox(0,0)[b]{$1$}}
\put(222,618){\makebox(0,0)[b]{$0$}}
\end{picture}
\end{center}
\caption{A strategy for the function of Example
1.}\label{ins}
\end{figure}
One can also compute the expected cost by summing up the product of the probability of ending up at each leaf and the cost of the path from the root to that leaf. Alternatively, one can try to compute the probability that the value of each variable is learnt and compute the expected cost over all variables.

The size of a $BDT$ can be exponential in terms of the input size. On the other hand, in order to execute a \emph{strategy}, we do not need to know the whole tree explicitly. It suffices to have an efficient algorithm that outputs the next variable to learn or the correct value of the function $f$, given the values of the variables learned so far. By running this algorithm at most $n$ times, we can learn the correct value of the function at any  $x \in \{0,1\}^n$

In general, in an optimal strategy, the next variable to learn  may depend on the results of previously conducted tests. We can refer these as \textit{adaptive strategies}. On the other hand, it is also a relevant problem to find a best strategy among those strategies where the next variable to learn does not depend on the values  of the previously learned variables  (as long as the variable is relevant in determining the value of the function). These strategies can be described more practically and efficiently as a permutation. We can refer to such strategies as \textit{nonadaptive strategies} or \textit{permutation strategies}. The ratio of the optimal expected cost of nonadaptive strategies over the optimal expected cost of adaptive strategies is defined as the \textit{adaptivity gap}. Essentially, adaptivity gap measures how much we lose, percentage-wise, in terms of the expected cost if we confine ourselves to only nonadaptive strategies at the expense of being able to describe the strategy concisely. On the other hand, there are some cases where there exists an optimal strategy that is nonadaptive or some algorithms by construction may produce nonadaptive strategies. 

\section{Classes of Monotone Boolean Functions\label{classes}}

A Boolean function $f:\{0,1\}^n \mapsto \{0,1\}$ is called {\em monotone} if $f(\bx)\geq f(\by)$ whenever $\bx \geq \by$, where $ \mathbf{x} \geq \mathbf{y}$  means that $x_i \geq y_i$ for all $ i \in \{1, \ldots, n\} $.

A monotone Boolean function can be represented by its unique, minimal  Disjunctive Normal Form (DNF) $f(\bx)=\bigvee_{I\in\cI(f)}\left(\bigwedge_{x_j\in I}x_j\right)$, where $I$ represents the implicants (terms) of and $\cI(f)$ is the set of all implicants of $f$.  Similarly, the dual of a monotone Boolean function can be represented by its unique minimal Conjunctive Normal Form (CNF) $f^d(\bx)=\bigwedge_{J\in\cJ(f)}\left(\bigvee_{j\in
J}x_j\right)$ where $J$ represents the implicates (clauses) of $f$ and $\cJ(f)$ is the set of all implicates of $f$.

In many papers and applications, the function of interest is a monotone Boolean function or assumed to be a monotone Boolean function. For example, for applications where the system can be modeled as a reliability system, changing the value of one of the variables from 0 to 1 typically means fixing the functionality of some component. So in these types of applications, the underlying function is naturally monotone. From a theoretical point of view, if the function is not monotone, it is a difficult problem to output the correct value of the function, when the value of the function can be determined by using  the values of some of the  variables. We now describe some interesting subclasses of monotone Boolean functions.
\subsection{Simple Series and Parallel
Systems\label{seriesparallel}}

A series
function (AND function, conjunction) takes value 0 if any one of its variables takes value 0, whereas a parallel
function (OR function, disjunction) takes value 1 if any one of its variables take value 1.  The series and parallel functions can be
written (respectively) in the following way.

$$f({\bf x}) = x_1 \wedge x_2 \wedge... \wedge x_n \mbox{ and } f({\bf x})=
 x_1 \vee x_2 \vee ...\vee x_n .$$

In spite of their simplicity, series and parallel functions arise
in numerous applications. In addition, all other known solvable
cases are generalizations of series and parallel functions. It is shown in \cite{But72} that it is optimal to test the variables in non-decreasing order of $c_i/q_i$ for a series function and in non-decreasing order of  $c_i/p_i$ for a parallel function. In other words, if 
$\frac{c_{\pi(1)}}{q_{\pi(1)}} \leq \frac{c_{\pi(2)}}{q_{\pi(2)}} \leq \ldots \leq \frac{c_{\pi(n)}}{q_{\pi(n)}}$ 
and 
$\frac{c_{\sigma(1)}}{p_{\sigma(1)}} \leq \frac{c_{\sigma(2)}}{p_{\sigma(2)}}\leq \ldots \leq \frac{c_{\sigma(n)}}{p_{\sigma(n)}}$ then it is optimal to test the variables in 
$\pi$ order for a series function and in $\sigma$ order for a parallel function.

\subsection{Threshold and $k$-out-of-$n$ Functions\label{kofn}}

A Boolean function $f:B^n\mapsto B$ is called {\em threshold} if there exists non-negative weights $w_1$, $w_2$, $...$, $w_n$ and a constant $t$ such that

\[
f(\bx) = \left\{
\begin{array}{rl}
1& \sum_{i=1}^nw_ix_i \geq t\\
0& \mbox{otherwise.}
\end{array} \right.
\]

A $k$-out-of-$n$ system functions if at least $k$ of its $n$
components function. In particular, a series system is an
$n$-out-of-$n$ system and a parallel system is a $1$-out-of-$n$
system. A $k$-out-of-$n$ function is a threshold function with equal weights and $t=k$. Generalizing the result for series (parallel) functions, polynomial-time optimal algorithms are provided in \cite{Dov81,CSF90} for $k$-out-of-$n$ functions. It turns out that learning next a variable that is in the first $k$ elements with respect to $\pi$ and the first $n-k+1$ elements with respect to $\sigma$ provides an optimal strategy.


First, we define {\em Regular Systems} before describing {\em Double
Regular Systems}. Given a monotone Boolean function
$f:B^n\mapsto B$, we say that the variable $x_i$ is {\em
stronger} than $x_j$ ($i\neq j$) with respect to $f$, and write
$x_i \succeq_f x_j$, if for all binary vectors $\bx\in B^n$ with
$x_i=x_j=0$ we have $f(\bx \vee \be_i)$ $\geq$ $f(\bx\vee \be_j)$,
where $\be_i$ denotes the unit $i^{th}$ unit vector.  If $x_i
\succeq_f x_j$ and $x_j \succeq_f x_i$ we shall say that $x_i$ and
$x_j$ are {\em equivalent with respect to} $f$ and write $x_i
\approx_f x_j$. 
Given a permutation
$\rho$ of  $\{ 1,2,...,n\}$, a Boolean function $f$ is
called $\rho$-{\em regular}, if its strength pre-order is
consistent with $\rho$, i.e. if $x_{\rho_1}$ $\succeq_f$
$x_{\rho_2}$ $\succeq_f$ $\cdots$ $\succeq_f$ $x_{\rho_n}$.  A
function is simply said to be {\em regular}, if it is
$\rho$-regular for some permutation $\rho$. 

Threshold functions are regular. In particular, if  $f$ is
threshold with weights $w_1 \geq w_2 \geq ...\geq w_n$, then $f$
is regular with respect to $\rho=(1,2,...,n)$. Since a
$k$-out-of-$n$ function is threshold with equal weights, it is
regular with respect to any permutation.

We shall say that a function is {\em Double Regular} if it is
$\sigma$-regular and $\pi$-regular for permutations $\sigma$ and
$\pi$.  The result for $k$-out-of-$n$ functions was generalized in \cite{boros1999diagnosing} by providing an optimal algorithm for double-regular functions with respect to $\sigma$ and $\pi$, as defined before.




\subsection{Read-Once Functions \label{SPS}}

Another interesting class of Boolean functions is the class
of {\em Read-Once functions} (also called
{\em Series-Parallel Functions}). Essentially Read-Once functions describe the connectivity between two specific nodes, the source and the sink,  in a specially structured network, called Series-Parallel networks (SPN). Each variable of the function describes the the functionality of a node in the network . The function takes
the value $1$ if there is a path between the source and the sink consisting of functioning nodes. The simplest
SPN consists of only one component: a single link connecting the
source to the sink. All other SPNs can be defined recursively as
a series or parallel connection of smaller SPNs. A {\em series
connection} identifies the sink of one SPN with the source of another, while a {\em parallel connection} identifies the
sources and sinks of the two systems. Formally, if
$f_1$ and $f_2$ are two functions with no
common variables, then $f_1\vee f_2$ is
their parallel connection, and $f_1\wedge
f_2$ is their series connection. All smaller SPNs, appearing recursively in the above
definition, are called the {\em subfunctions} of the considered SPN.
It is possible to describe the connectivity of the source and the sink as a Boolean formula where each variable appears only once for these functions; hence the name Read-Once functions.

For example, the network in figure \ref{tonguc hoca1}(a)
is an SPN.  We can  represent the
associated function by a logical expression as follows:

\begin{equation}\label{cr4e0}
f(x_1,x_2,x_3,x_4)=x_1\vee \left( x_2 \wedge \left( x_3 \vee
x_4\right)\right) .
\end{equation}

The network in figure \ref{tonguc hoca2}, is not an SPN, since it cannot be obtained by the recursive procedure described above, while the first one is an SPN, formed by a parallel connection of
the two subsystems corresponding to variables $\{ x_1\}$ and $\{x_2,x_3,x_4\}$.  The latter
one itself is a series connection of two smaller subsystems, $\{
x_2\}$ and $\{ x_3,x_4\}$, where this last subsystem is a simple parallel network.  


\begin{center}
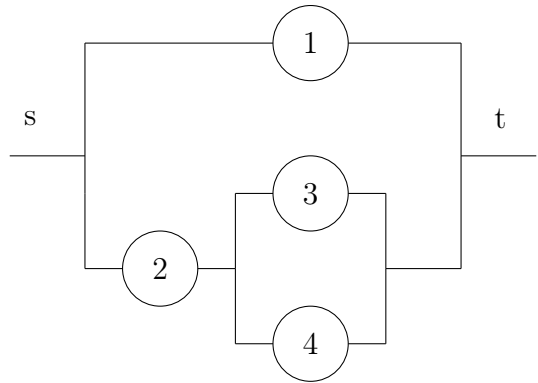

\begin{tikzpicture}
    \draw (5,4) circle [radius=0.5] node {1};
    \draw (3,1) circle [radius=0.5] node {2};
    \draw (5,2) circle [radius=0.5] node {3};
    \draw (5,0) circle [radius=0.5] node {4};
    
    \draw (1,2.5) -- (2,2.5);
    \draw (2,2) -- (2,4);
    \draw (2,4) -- (4.5,4);
    \draw (2,2) -- (2,1);
    \draw (2,1) -- (2.5,1);
    \draw (3.5,1) -- (4,1);
    \draw (4,1) -- (4,2);
    \draw (4,1) -- (4,0);
     \draw (4,0) -- (4.5,0);
    \draw (4,2) -- (4.5,2);
     \draw (5.5,0) -- (6,0);
     \draw (5.5,2) -- (6,2);
     \draw (6,1) -- (6,2);
     \draw (6,0) -- (6,1);
     \draw (6,1) -- (7,1);
      \draw (5.5,4) -- (7,4);
      \draw (7,4) -- (7,2.5);
       \draw (7,1) -- (7,2.5);
       \draw (7,2.5) -- (8,2.5);

   \draw (1.5,3) node[left] {s};
    \draw (7.75,3) node[left] {t};
    

\end{tikzpicture}
     \captionof{figure}{Series-Parallel Network }
     \label{tonguc hoca1}
\end{center}
\begin{center}
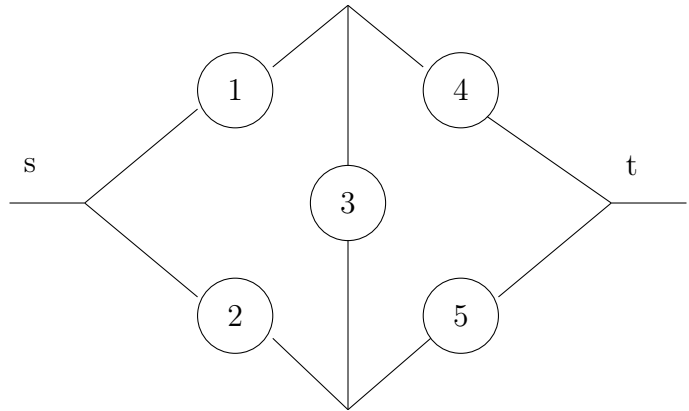

\begin{tikzpicture}
    \draw (3,4.5) circle [radius=0.5] node {1};
    \draw (3,1.5) circle [radius=0.5] node {2};
     \draw (4.5,3) circle [radius=0.5] node {3};
     \draw (6,4.5) circle [radius=0.5] node {4};
    \draw (6,1.5) circle [radius=0.5] node {5};
    
    \draw (0,3) -- (1,3);
    \draw (1,3) -- (2.5,4.25);
    \draw (1,3) -- (2.5,1.75);
    \draw (3.5,4.81) -- (4.5,5.63);
    \draw (5.5,4.81) -- (4.5,5.63);
     \draw (4.5,3.5) -- (4.5,5.63);
    \draw (4.5,2.5) -- (4.5,0.25);
    \draw (3.5,1.20) -- (4.5,0.25);
     \draw (4.5,0.25) -- (5.6,1.20);
     \draw (8,3) -- (6.5,1.75);
      \draw (8,3) -- (6.35,4.15);
       \draw (8,3) -- (9,3);
      \draw (.5,3.5) node[left] {s};
    \draw (8.5,3.5) node[left] {t};

    

\end{tikzpicture}
     \captionof{figure}{A network that is not a Series-Parallel Network }
     \label{tonguc hoca2}
\end{center}

The {\em depth} of an SPN is a measure of the number of series
and/or parallel combinations in the system and is defined as
follows: The depth of a simple series or a simple parallel system
is $1$. The depth of any other SPS is $1 + \max \{ \mbox{depth of
proper subsystems}\}$. For example, the SPN in figure
\ref{tonguc hoca1} (a) has depth $3$. Thus, every SPN is either a
parallel or series connection of another SPN whose depths are at
least $1$ lower than the original SPN. We note that the Boolean dual of a read-once function is another read-once function with the same depth that can be obtained by interchanging $\vee$ and $\wedge$ in the read-once description.

Optimal algorithms are provided for 2-level Series Parallel functions and 3-level Series Parallel functions with identical variables (i.e. the costs and probabilities are the same for all variables) in \cite{boros2000sequential}. These algorithms essentially start at the deepest level where we have a series or parallel subsystem, and tests these subsystems optimally. One may refer to these algorithms as \textit{depth-first algorithms}. Depth-first algorithms are also discussed in \cite{greiner2006finding,Dov81}. 

We would like to define some closely related problems that have been studied in the literature. The \textit{stochastic score classification problem, (SSCP)} was introduced in \cite{gkenosis2018stochastic}. In this problem, each patient belongs to one of the finite number of classes that depends on the number of tests with positive results, where each test result is either negative or positive. The goal is to find out the class of the patients by conducting the minimum expected cost or number of tests. SSCP generalizes STP for $k$-out-of-$n$ functions. In particular STP for $k$-out-of-$n$ functions is an SSEP with two classes.  In the \textit{stochastic half-space evaluation problem (SHEP)} introduced in \cite{ghuge2024nonadaptive}, the goal is to learn whether a binary vector satisfies a number of binary linear inequalities or not. In other words, if we define a function that gets a value 1 if all the binary linear inequalities hold and 0 if at least one of the binary linear inequalities do not hold, the goal in this problem is to learn the value of this function.
Another variation of the problem has been introduced in \cite{daldal2017sequential} where it is allowed to execute a subset of tests at once for the series function. In order words, it is possible to learn the values of the variables as a \textit{batch}.  In this case, the cost structure is slightly different and for each subset of tests conducted, there is a fixed cost and a variable cost for each test in the subset. This batch testing concept can also be incorporated for other functions for STP as well as  SSCP and SHEP.

\section{Models, Applications, Solution Methods and Results}\label{rev}

In this section, we provide a review of the studies in terms of application areas, type of solution approach, the assumptions of the models, and the function. So, a paper can be related to more than one of these different categories, and the results can also be interrelated. The papers are from a wide variety of applications with different motivations. Consequently, at this breadth level, it is not possible to come up with a taxonomic categorization of the papers. 

\subsection{Function Evaluation}\label{evaluation}

The STP has been studied for different classes of functions and algorithms have been developed by exploiting the special structure of the functions. In addition, researchers have also considered special cases of data, in particular the case of all tests having the same cost and/or probability.  Numerous hardness results are provided along with approximation algorithms that have been proposed for different classes of functions. In this section, we review these results. First, we will discuss the papers that present approximation algorithms and hardness results.

In \cite{kaplan2005learning}, it is shown that when the test results follow a general distribution function (the tests are not necessarily independent), the problem is NP-complete for a parallel (series) function, and a 4-approximation algorithm is provided.  This result is extended for 2-level read-once functions. General monotone functions where the cost of learning any variable is unity and the variables are drawn from a uniform distribution (so each variable takes value 1 is 1/2) are also considered. Learning a variable that appears in the maximum number of implicants and the maximum number of implicates is shown to result in a strategy whose expected cost is at most $O([E_D|CER(f,x)]\ln m)$ where $CER(f,x)$ is the cheapest implicant or implicate of $f$. The cost of an implicant or implicate is measured as the sum of the testing costs of the variables of the implicant or implicate. Since costs are unity in this case, the cost of an implicant(implicate) is the number of variables in the implicant(implicate).   Here, $m$ is the maximum of the number of implicants  in the DNF and implicates in the CNF and  the expectation taken over $D$, the underlying uniform distribution. This result is significant for functions with small DNF/CNF. In this paper, a round-robin approach is utilized for the approximation algorithms. Modified versions of this approach have been used in other papers. One intuitive reason why such a method is useful is the following. The main goal is to show that the function is 0 or 1 and in certain cases two different ideas may serve well to show that the function is 0 or 1. Using these in an algorithm in turn can lead to approximation results.

One particular line of research has utilized results from adaptive submodularity to develop approximation algorithms. The approach referred to as the Q-value approach essentially reduces the problem to an instance of Stochastic Submodular Set Cover (SSSC) and then solves it using the Adaptive Greedy algorithm of \cite{golovin2011adaptive}. One of the critical steps in this approach is to define an appropriate Q-value that will work for the particular class of functions studied.

In \cite{deshpande2014approximation,deshpande2016approximation}, an algorithm called adaptive dual greedy is developed for the SSSC problem and this is used to obtain a 3-approximation algorithm for the STP for threshold functions. The Q-value approach also yields an $O(\log kd)$-approximation algorithm for the case DNF and CNF representations are given for a monotone Boolean function where $k$ is the number of clauses in the CNF and $d$ is the number of terms in the DNF. 

In \cite{allen2017}, $k$-DNFs and $k$-term DNFs with arbitrary cost  are considered. A $k$-term DNF formula consists of at most $k$ terms; a $k$-DNF formula
consists of terms containing at most $k$ literals. For a $k$-term DNF, an approximation algorithm that is within $\max\{2k,\frac{2}{\rho}(1+\ln k)\}$ is provided whereas  for a $k$-DNF an approximation algorithm that is within $4/\rho^k$ times the optimal value where $\rho$ is $\min_i\{p_i,1-p_i\}$ is provided.
 In \cite{gkenosis2022stochastic}, a $O(\log n)$ approximation algorithm for symmetric Boolean functions (these are Boolean functions for which the value of the function can be determined by the number of variables that are equal to 1) using the goal value approach of \cite{deshpande2016approximation} is provided. Symmetric Boolean function generalizes $k$-out-of-$n$ functions. Symmetric Boolean functions can be represented by a value vector of size $n+1$ where the $j^{th}$ value of this vector is the value of $f$ if there are exactly $j$ variables that are equal to 1. The approximation factor of the second algorithm is $B-1$ where B is the number of blocks of 0's and 1's in the value vector for $f$.

 In \cite{blanc2021query}, a general Boolean function is considered where the values of the variables are uniform (i.e. the variables take value 0 or 1 with probability 0.5). It is shown that there is a $O(\epsilon)$ randomized strategy whose expected cost is at most $Inf_f/\epsilon^2$ times the optimal value where $\epsilon$ is an error parameter and $Inf_f$ is the sum of the influence of the variables. The influence of a variable $x_i$ is defined as the probability that the value of $f$ changes if the value of variable  is switched. The algorithm randomly tests a variable with the highest $Inf/c_j$ value.

 In \cite{happach2022general}, an 8-approximation algorithm for the nonadaptive problem for read-once functions is presented.  The 8-approximation factor that runs in pseudopolynomial time in terms of the sum of the testing costs, is with respect to the optimal nonadaptive strategy, and not with respect to the optimal adaptive strategy. In \cite{hellerstein2022adaptivity} it is shown that the adaptivity gap for evaluating read-once DNF is $\Theta(\log n)$ for unit costs and uniform distribution, so there is a big difference between the optimal adaptive and the optimal nonadaptive strategies. In addition, in \cite{hellerstein2022adaptivity}, the authors study the adaptivity gap for DNFs  and read-once formulas and report relatively negative results. For DNFs the adaptivity gap is $\Omega(n/ \log n$ for unit costs and uniform distribution, and for read-once formulas the adaptivity gap is $\Omega(\epsilon^3n^(1-2\epsilon/\log 2)$ for any fixed $\epsilon>0$ for unit costs and equal probabilities.

There are some other papers presenting exact and heuristic methods for some cases, which we discuss next.
In \cite{greiner2002optimal}, the STP is studied for read-once functions. It is shown that the best depth first strategy is optimal for 1 or 2 level series-parallel systems consistent with \cite{boros2000sequential}. For more complicated read-once functions, negative results are reported. For instance, for general read once functions there are other nonadaptive strategies that are arbitrarily better than the strategy produced by the depth first algorithm. In \cite{greiner2006finding}, a dynamic programming approach is proposed to solve the problem for read-once functions with complexity $O(d^2(d+1)^r)$ where $d$ is the number of leaf-parents and $r$ is the largest number of leaf-siblings
In \cite{azaiez2007optimal}, the STP is considered from a different perspective. An attacker is trying to destroy a system, and the defender is trying to allocate resources to different components of the system to make them less vulnerable. The resulting problem turns out to be the STP for read-once functions. The authors incorrectly claim an optimal algorithm for read-once functions (representing series-parallel networks). The algorithm is essentially the depth-first strategy of \cite{greiner2002optimal} and the SPS Permutation algorithm of \cite{boros2000sequential}. It turns out that this algorithm can produce arbitrarily bad solutions, as described in \cite{unluyurt2009note}. In \cite{unluyurt2005testing}, the STP with identical variables is considered. In other words, all testing costs and probabilities are assumed to be the same for all variables. The notion of lexicographically large binary decision trees is introduced, and a heuristic algorithm based on that notion is proposed. The performance of the heuristic algorithm is demonstrated by computational results for various classes of functions, including threshold functions.\\

As defined in section \ref{classes}, the stochastic score evaluation problem was introduced in \cite{gkenosis2018stochastic}.
This problem generalizes the STP for threshold functions and $k$-out-of-$n$ functions. Essentially, this is a function that takes discrete values where the value of the function depends on the number of variables having value 1.
In \cite{grammel2022algorithms},  a polynomial time 2-approximation algorithm is provided for the adaptive and nonadaptive case for the unit cost case. In \cite{plank2024simple}, a simple nonadaptive approximation algorithm for the score classification problem that achieves an approximation factor of $3 + 2\sqrt{2} \approx 5.828$ for the general cost case. In addition, using an idea previously studied in \cite{gkenosis2018stochastic}, it is possible to obtain only slightly larger guarantee of 6. Similarly in  \cite{ghuge2024nonadaptive}, a constant factor approximation algorithm is provided for the SSCP for general costs. In this work, the SHEP is studied generalizing the STP of threshold function (STP for threshold functions is SHEP with a single half-space). In the stochastic half-space evaluation problem, the goal is to learn whether a binary vector satisfies a number of binary linear inequalities or not. In other words, if we define a function that gets a value 1 if all the binary linear inequalities hold and 0 if at least one of the binary linear inequalities do not hold, the goal in this problem is to learn the value of this function. An approximation algorithm is provided for this more general problem with a factor that depends on the number of half spaces, that is $O(d^2 \log d)$, where $d$ is the number of half spaces.

In \cite{hellerstein2024quickly}, an election is considered in which $n$ voters can vote for $d$ candidates. It is assumed that $p_{ij}$ is the probability that the voter $i$ chooses the candidate $j$, independent of other voters. After the voting is complete, it is possible to learn the vote of any voter in a certain time that depends on the voter. The goal is to solve the problem of determining the results in the minimum expected time.  Polynomial-time constant factor approximation algorithms for both the absolute-majority (approximation factor of 4) and the relative-majority (approximation factor of 8) versions are provided.

In Table \ref{tab:table1}, the important and general results are summarized for STP, SSCP and SHEP. In the first two columns, the related article is mentioned, and the problem studied is described. In the third and fourth columns, the information about the data is given. For probabilities, "product" means that the variables take values independently of each other. The last column describes the main results.

 \begin{table}[ht!]
\centering
\caption{Summary of results\label{tab:table1}}
\begin{tabular}{|>{\raggedright\arraybackslash}p{3cm}|>{\raggedright\arraybackslash}p{3cm}|>{\raggedright\arraybackslash}p{3cm}|>{\raggedright\arraybackslash}p{3cm}|>
{\raggedright\arraybackslash}p{3cm}|}
\hline
\textbf{Article} & \textbf{Type of Problem-Function} & \textbf{Probabilities} & \textbf{Costs} &\textbf{Result} \\
\hline
\cite{boros1999diagnosing} & STP-double regular & Product & General & Exact\\
\hline
\cite{boros2000sequential} & STP-Series-Parallel (upto 3 levels)  & Product & General & Exact\\
\hline
\cite{kaplan2005learning} & STP-Series & General & General & 4-Approx. \\
\hline
\cite{deshpande2014approximation,deshpande2016approximation} & STP-Threshold & Product & General& 3-Approx.\\
\hline
\cite{deshpande2014approximation,deshpande2016approximation} & STP-DNF, CNF & Product & General& $O(\log kd)$-Approx.\\
\hline
\cite{happach2022general} & STP Series-Parallel & Product & General & 8-Approx. (nonadaptive)\\
\hline
\cite{allen2017}& STP-k-DNF & Product & General & 1/$\rho^k$-Approx.\\
\hline
\cite{allen2017}& STP-k-term DNF& Product & General & 2k-Approx.\\
\hline
\cite{grammel2022algorithms} & SSCP & Product & Unit & 2-Approx.\\
\hline
\cite{plank2024simple} & SSCP & Product & General & 5.83-Approx.\\
\hline
\cite{ghuge2024nonadaptive} & SSCP &  Product & General  & Const.-Approx.\\
\hline
\cite{ghuge2024nonadaptive} & SHEP &  Product & General  &$ O(d^2 \log d)$-Approx.\\
\hline
\end{tabular}
\end{table}

\subsection{Batch testing} \label{batch}

 The structure of the problem is discussed and heuristic algorithms are proposed in \cite{daldal2017sequential}.  Later approximation algorithms for the case where any subset of all subsets can be batched have been developed. A polynomial time approximation algorithm with an approximation factor of $6.929+\epsilon$ for any fixed $\epsilon \in (0,1)$ is developed in \cite{daldal2016approximation}. Later the approximation factor is improved to $1+\epsilon$ for any $\epsilon>0$ in \cite{segev2022polynomial} by developing and leveraging a number of techniques in approximate dynamic programming. In \cite{ghuge2024nonadaptive}, a constant factor approximation algorithm is provided for the SSCP generalizing the series system. In both cases, the batch-cost structure is assumed to be as described in \cite{daldal2017sequential}.

In \cite{tan2024general}, an approximation algorithm for batched sequential testing (for any system) that leverages any nonadaptive approximation algorithm for the classical STP. The approximation ratio for the batched setting is only a small constant (at most 0.71) more than the classical (unbatched) setting. Combined with previously known approximation algorithms in the classic setting, small constant-factor approximation algorithms for batched STP for series functions (, $k$-out-of-$n$ functions and  the SSCP are obtained. The approximation factors are 1.707, 2.268 and 6.371, respectively. The approximation factor for the SSCP with unit costs is 2.618. The algorithms run in strongly polynomial time.  The results for the case where the values of the variables can be learned in batches are summarized in Table \ref{tab:table2}. The organization of the table is the same as Table \ref{tab:table1}.

 \begin{table}[h!]
\centering
\caption{Summary of results for the batch testing case \label{tab:table2}}
\begin{tabular}{|>{\raggedright\arraybackslash}p{3cm}|>{\raggedright\arraybackslash}p{3cm}|>{\raggedright\arraybackslash}p{3cm}|>{\raggedright\arraybackslash}p{3cm}|>
{\raggedright\arraybackslash}p{3cm}|}
\hline
\textbf{Article} & \textbf{Type of Problem-Function} & \textbf{Probabilities} & \textbf{Costs} &\textbf{Result} \\
\hline
\cite{daldal2016approximation} & STP-series & Product & General & 6.829-Approx\\
\hline
\cite{segev2022polynomial} & STP-Series & Product & General & (1+$\epsilon)$-Approx.\\
\hline
\cite{ghuge2024nonadaptive} & SSCP & Product & General & Const.-Approx.\\
\hline
\cite{tan2024general} & SSCP & Product & General & 5.371-Approx.\\
\hline
\cite{tan2024general} & SSCP & Product & Unit Cost & 2.618-Approx.\\
\hline
\cite{tan2024general} & STP-Series & Product & General & 1.707-Approx.\\
\hline
\cite{tan2024general} & STP-$k$-out-of-$n$ & Product & General & 2.268-Approx.\\
\hline
\end{tabular}
\end{table}

In \cite{yang2024sequential}, the batch testing of a series system is considered with an additional resource constraint. The problem is formulated as a mixed integer program and a dynamic programming approach is proposed to obtain an exact solution. For larger problem instances tabu search, a hybrid heuristic method that combines a tabu search metaheuristic and a proximity search matheuristic are proposed.

\subsection{Precedence Constraints}

As mentioned in section \ref{intro}, in some applications there could be precedence constraints among tests due to logical, physical, or technological reasons. The precedence constraints can be naturally described as a directed cyclic graph, where the nodes correspond to variables, and an arc from node $i$ to node $j$ means that one can learn the value of $x_j$ after learning the value of $x_i$. In this line of research, typically, a certain class of functions with a certain type of precedence constraints is studied. An early result for series functions under precedence constraints is a polynomial algorithm when the precedence graph is a collection of out-trees, \cite{Gar73}. This is an elegant algorithm that uses conditions under which two variables should be learned one after another and conditions that allow the removal of an arc in the precedence graph.

One particular line of research where precedence constraints arise has been motivated by an application in project management. In this context, it is assumed that a project consists of tasks that have to be executed sequentially, and the payoff occurs when all of these tasks are completed successfully.  Each task has an associated probability of being successfully executed. This framework is quite similar to the STP where the function under consideration is a series system, and there are precedence constraints with discounting. In \cite{de2008r} the problem of determining the start time of each task while maximizing the net present value for the project is considered. The problem is shown to be NP-hard and a branch-and-bound algorithm is proposed to solve the problem. 

In \cite{bao2008optimal}, STP for series systems under block-type precedence constraints is considered. In this case, in the precedence graph, the in-degree and out-degree of each node is at most 1. An optimal algorithm is provided for this special type of precedence graph. Actually, block-type precedence constraints is a special case of the precedence graphs considered in \cite{gabbert2013sequential}. It is claimed that under certain conditions the same idea provides optimal solutions for some $k$-out-of-$n$ functions.

An ant colony algorithm is proposed in \cite{ccatay2011testant} for the STP for a series series for general precedence constraints and evaluated against a greedy algorithm and the optimal solution for small-sized problems over randomly generated problem instances.

In \cite{rostami2019sequential}, STP for series systems under precedence constraints is considered. A branch and price and a dynamic programming based methods are proposed to obtain exact solutions by strengthening the precedence graph. In \cite{berend2014optimal}, it is shown that the STP for a simple series function with general precedence constraints is NP-complete. For certain type of precedence constraints and data, it is shown that certain tests are executed one after another in an optimal solution. The basic logic of this coincides with the results in \cite{rostami2019sequential}. In \cite{wei2013sequential,leus2012sequential}, STP for $k$-out-of-$n$ functions under precedence constraints is considered and two exact methods, namely, one branch and bound algorithm and one dynamic programming algorithm are proposed. The effectiveness of the algorithms are demonstrated using randomly generated data.

\subsection{Network Connectivity}

Another application area has been determining some kind of connectivity in networks. In this case, the function describes the defined connectivity and the goal is to verify whether the network is connected or not. Typical application areas are telecommunications networks, road networks, post-disaster operations. The read-once functions that were mentioned before also describe the connectivity of a special type of network, namely series-parallel networks. We summarized the results related to read-once functions in section \ref{evaluation}. In this section, we review the results for more general networks. 

In \cite{fu2014optimal,fu2016we}, the problem of determining connectivity of two designated  nodes in Erdös-Renyi random graphs where an edge exists between two nodes with probability $p$ is considered. The testing costs are assumed to be the same for all arcs. It is shown that the next arc to test is on a shortest path between the designated nodes and a minimum cut disconnecting the designated nodes. The result can be extended to some more general graphs, and these results coincide with known results for read-once  functions with identical variables in terms of costs and probabilities.

In \cite{fu2017determining} the network connectivity problem is studied between a pair of nodes in a general network for arbitrary costs and probabilities. In this case, the cost of testing an edge is constant and does not depend on the previous edge tested. It is shown that the problem is NP-complete. Markov decision process and dynamic programming based exact solution algorithms are proposed, as well as an approximation algorithm using greedy adaptive submodularity.

In \cite{guo2024limited}, the network connectivity problem is studied again when the testing costs are the same for each edge and each edge appears with the same probability. In addition, there is a given upper bound on the number of tests and when that upper bound is reached, testing stops without conluding about the connectivity. The problem is motivated by a cyber-security use case where the goal is to figure out whether attack paths exist in a given network. Edge query is resolved by manual effort from the IT admin. The problem is shown to be $\#P$ hard.  An empirically scalable exact algorithm along with heuristic algorithms based on reinforcement learning and Monte Carlo Tree search are proposed to solve the problem. 

In these studies, the cost structure is the same as the standard STP. Yet for checking the connectivity of a physical network, one of the fundamental assumptions of the problem may change to give rise to an even more challenging problem. The costs may depend on the previous test executed, since one has to travel from one node in the network to another to execute the next test and incur some costs.   We have not found any references for such a problem in the literature for the network connectivity case. Yet some similar search problems (e.g. \cite{teller2019minimizing, khare2023improving} that will be reviewed in section \ref{other} consider a cost structure where the cost of the next test is dependent on the previous test.

\subsection{General Discrete Functions}\label{general}

In some studies, general discrete functions are considered as opposed to Boolean functions. The variables typically take Boolean values (in the most general case values from a finite discrete set) and the function can assume a value from a finite discrete set. Essentially, the values that the function can assume correspond to different states of the system that the function represents. For instance, the different states may correspond to distinct failure states. This can also be considered as a sample version of the sequential version of the problem since the input in this case is a subset of $\{0,1\}^n$ and the value of the function for each element of this subset.
The goal still is  to find a strategy that gives the minimum expected cost of the correct state of the system for the given sample. In this line of research, typically heuristic algorithms are proposed and these algorithms are tested on problem instances derived randomly or based on some real system.

In \cite{tu2003rollout} an exact dynamic programming approach and  rollout strategies that utilize concepts from information theory, Huffman coding,  heuristic AND/OR graph search methods are proposed.  The methods are tested using randomly generated data in addition to 11 instances obtained from real engineering systems. It is demonstrated that the information heuristic algorithm outperforms other approaches.
In \cite{kundakcioglu2007bottom}, a bottom-up approach (as opposed to many top down approaches) is proposed by using look-ahead heuristic algorithms and basic ideas of Huffman coding. The effectiveness of the algorithm is demonstrated in problem instances of different sizes. In \cite{yang2014novel}   a self-adaptive discrete particle swarm optimization algorithm is proposed whereas in \cite{wang2022general},  a hybrid heuristic based on support vector machines, evolutionary clustering algorithm and Monte Carlo simulation is proposed. In \cite{zhang2023divide}, a divide and conquer algorithm using information entropy is proposed. In all of these studies, the effectiveness of the algorithms is demonstrated using randomly generated data and/or real data available from the literature for some mechanical/electrical systems.

In \cite{feldman2010model},  an approach for reducing the diagnostic uncertainty, called active testing, is developed. This approach generalizes sequential diagnosis and model-based diagnosis, allowing  combination of multiple passive sensor readings, and does not require explicit tests and test dictionaries. Various practical algorithms are proposed and tested using  empirical data on 74XXX/ISCAS85 circuits.

In \cite{bellala2012group}, the problem is studied from different perspectives, where objects are partitioned into groups, queries are partitioned into groups, and the case with query noise. Algorithms are proposed to handle all these different perspectives and tested on generated data and a database used for toxic chemical identification.

The only approximation algorithm for this version of the problem has been proposed in 
\cite{cicalese2017decision,cicalese2014diagnosis}. The minimization of both the expected cost and the worst-case cost of a decision tree are considered simultaneously for this sample version of the problem. $\log n$-approximation algorithms are proposed for the sample version of the problem. 

\subsection{Imperfect tests} \label{imperfect}

There are also studies that consider imperfect tests which is a realistic assumption in many applications. In this case, the problem structure is rather different yet there are important commonalities with the STP. The structure of a strategy is the same as well as the computation of the expected cost of a strategy. Yet in this case, it is possible that the output is not correct. So this has to be taken into consideration by introducing a misclassification cost or an additional constraint on the probability of an incorrect output. Another alternative is to maximize the probability of a correct output given a certain budget for the inspection.

For instance in \cite{wei2017minimum}  $k$-out-of-$n$ functions are considered when the tests are imperfect. It is assumed that both type I and type II errors are possible. Th goal is to find minimum expected cost strategies while respecting a bound for the probability of incorrect output. So testing continues until a certain confidence level is reached or all components are tested. Under some technical assumptions, the case with imperfect tests can be converted to a  generalized testing problem with perfect tests and by this way the problem can be solved optimally in polynomial time. In \cite{wei2017test}, imperfect tests are considered along with precedence constraints for $k$-out-of-$n$ functions. A tabu search algorithm is provided together with a simulation-based evaluation technique that incorporates importance sampling to find high-quality solutions within limited run times. Another possibility in this case is to allow repeating a test. For instance, in \cite{shahmoradi2018failure}, a variation of the problem is considered with imperfect tests and each test can be repeated once.  In \cite{liang2018novel}, imperfect tests are considered for the sample version of the problem. A Markov decision process based algorithm is tested on a case study about the suspension and launcher test on a certain type of missile.

\subsection{Finding the cause of failure}\label{cause}

Another line of research assumes that a system has failed (or the value of the function is 0), and considers the problem of efficiently finding the reason for this failure. In this case, the individual variables are no longer independent of each other. Still, this is a special type of dependency. As a side note, it would be a relevant and interesting problem to model the case where the variables do not assume values independently. It would also be important to describe this dependency in a practically relevant manner.

This problem was first introduced and studied in \cite{nachlas1990diagnostic} for series functions. In this case, $p_i$ is defined as the probability that component $i$ is the cause of failure. It is assumed that the variables represent the life times of some components and they follow a continuous distribution. Hence when the system fails, it is known that this is due to the failure of a single component.  For a series system, an optimal strategy is shown to be one that tests components in non-decreasing order of $\frac{c_i}{p_i}$. One difference is that there is no need to test the last component if the first $n-1$ components are all found to be functioning.  
In \cite{shahmoradi2018failure}, an extension of this model is studied again for a series system where the tests are not perfect and it is a possible decision to repeat a test at most once. So a strategy should also describe the repetition policy along with the sequence of the tests. There are also misclassification costs when the output of the algorithm is incorrect. Various heuristic algorithms including local greedy, simulated annealing, and genetic algorithms are used to solve the problem.   In \cite{yavuz2019exact}, a similar setting is considered for $k$-out-of-$n$ functions. A $k$-out-of-$n$ function has value 0 due to the fact that $n-k+1$ variables have taken 0 value (or $n-k+1$ components have failed). The goal is to find the subset of variables causing this failure. In this case, probabilities are associated with each subset of the components of cardinality $k$. Integer programming formulations are provided for this case and a Markov Decision Process based algorithm is proposed to solve the problem. In \cite{kovalyov2006optimal}, a series system that has just failed due to the failure of one of its components. It is possible to repair a component directly or test before repairing it.  They show that the problem of finding a sequence of repairs and tests to minimize total expected cost can be modeled as minimizing a quadratic pseudo-Boolean function. Polynomially solvable special cases of the latter
problem are identified and a fully polynomial time approximation scheme is derived for the general case.

In \cite{wang2009fault,wang2011fault}, a sequential testing
problem that combines active and passive measurements for fault localization in wireless sensor networks is formulated.
This problem determines an optimal testing sequence of network components based on end-to-end data to minimize testing cost.  A recursive approach and two
heuristic algorithms are proposed to solve the problem. Extensive simulation
demonstrated that heuristic algorithms only require a
few iterations and testing a small subset of components to
identify all lossy components in  quite special types of  networks.

\subsection{Efficient Querying in Databases}\label{database}

Another application area is designing efficient query answering strategies where data are stored in a distributed network. For instance,  in \cite{laber2004querying,carmo2007querying}, a query optimization problem is considered where  a set of tuples that satisfy certain conditions is sought for and checking these conditions. The overall goal is to minimize the user response time. In the simplest case, when the query looks for the set of tuples that simultaneously satisfy $k$ expensive predicates (that means the tuples that satisfy a set of conditions), the problem reduces to ordering the evaluation of these conditions so as to minimize the time to output the set of tuples comprising the answer to the query.  
In \cite{amsterdamer2019pepper}, a tool referred to as  PePPer is introduced  for fine grained, personal access control. The system enables loading data items from different sources and annotating them with fine-grained access control requirements
via provenance-style Boolean expressions. These Boolean expressions are
evaluated to decide whether the client is allowed to share the data
with a given peer, using a taxonomy that compactly captures data
ownership and access control policies. So, it is important to execute these evaluations efficiently.
In \cite{drien2021managing}, the problem of determining a strategy that minimizes the expected number of probes required to determine consent with respect to a query result is studied. The proposed framework considers the simultaneous evaluation of multiple, possibly many expressions corresponding to the
provenance of multiple tuples. In \cite{yang2015lenses}, the authors explore the design of a general extensible infrastructure for on-demand data curation that is based on probabilistic query processing. The process of selective parsing, transformation, and
loading into a new structure of data from different sources is referred to as data curation. Prioritizing curation tasks is quite similar to the STP and can be adaptive in general. Similarly in \cite{drien2023query}, the problem of query-guided uncertainty resolution is studied. Rather than verifying all input tuples to answer a query, they try to find a subset of the input tuples that suffices to determine the correct results of the given query efficiently.

In these studies, the general problem of efficiently evaluating Boolean functions is quite relevant. Yet, this is only one of the concerns, and their final goal is to develop an overall system that works practically. Sometimes the input of the problem is nor a monotone DNF or CNF or other special cases of monotone Boolean functions that have been studied. Still, there is potential in this area to implement the ideas from theoretical papers within an integrated overall system.

\subsection{Similar Models or Other Applications}\label{other}
In this section, we review the publications that are similar, but there are some fundamental differences from the typical framework or publications about different application areas. 

In \cite{charikar2002query}, a different version of the STP is considered where probabilistic information on the value of variables is not known. The goal is to evaluate the function by learning the values of the variables, which is costly. The strategy is still represented by a binary decision tree.  For a binary input vector whose individual entries are unknown, the performance metric is the ratio of the cost of the strategy incurred for that input over the least-cost proof for that input. The maximum ratio over all possible inputs is defined as the competitive ratio of the strategy. The overall goal is to find the strategy with the minimum competitive ratio.  It is shown that for every AND/OR tree, and every cost vector, the optimal competitive ratio can be achieved by an efficient
algorithm. Specifically, the algorithm has a running time that is polynomial in the size of the tree and the magnitudes of the costs, i.e. the algorithm is pseudo-polynomial. Similarly, in \cite{cicalese2011competitive,cicalese2006competitiveconf}, the problem of evaluating monotone Boolean functions is considered where there are only costs of learning variables but no probabilistic information. The goal is to find strategies with the minimum competitive ratio as in \cite{charikar2002query}. A polynomial time algorithm is proposed that has the best possible competitive ratio for each monotone Boolean function that is represented by a threshold tree. In addition, a linear programming approach is developed for handling the problem for monotone Boolean functions when the costs are not known in advance. 

\cite{chikalov2016totally} considers the problem of simultaneously optimizing different objectives related to binary decision trees representing monotone and arbitrary Boolean functions such as depth, average depth and number of nodes and proposes a dynamic programming based approach for this purpose. So, the costs are uniform, and there is no probability associated with the variables. The basic idea of this work is the existence of Boolean functions for which there  exist decision trees optimizing the mentioned measures.  In  \cite{saettler2017decision}, a similar problem as described in section \ref{general} is considered and among all nonadaptive strategies, the problem of finding  a minimum cost decision tree where all leaves are at the same depth and there is a bounded number of misclassifications is considered. An approximation algorithm that produces a  decision tree with an $O(\log n)$ approximation of the optimal cost and an error at a constant factor of $k$, where $k$ is the allowed number of errors.
In \cite{agnetis2022time}, a generalization of the STP is studied where the goal is to minimize the expected cost of testing while testing should stop before a deadline measured in terms of the number of tests executed. Another problem is also considered in this paper where multiple searchers try to find a target simultaneously. The problem is considered for a series system. It is shown that the problem is NP-hard. Heuristic search algorithms are proposed to solve the problem.
In \cite{kress2008optimal}, it is assumed that an object is hidden in one of the possible locations and the objective is to find the target as quickly as possible. The application mentioned is searching for a  hostage or detecting improvised explosive devises. Different cases of imperfectivity are considered for the tests, namely perfect, partial, or wrong detection. In \cite{lidbetter2019searching}, the problem of searching for multiple targets in possible locations is considered. A game theoretic solution method is proposed to solve the problem.  Search problems have been studied extensively in the thesis \cite{clarkson2020optimal}.

In \cite{teller2019minimizing},  a generalization of the STP has been studied where the cost of the next test to be executed depends on the previous test. This is a general model for applications where one has to travel physically to execute the tests. One example could be checking the status of infrastructure after a disaster using teams of personnel or drones. It is shown that the problem is NP-complete and dynamic programming and branch and bound based exact algorithms are proposed along with heuristic algorithms.

 In \cite{khare2023improving}, the problem of searching for gasoline after a disaster happens is modeled as finding an object on a graph with the minimum expected cost. In fact, this is part of a more general problem, and optimizing the search effort is the last stage of this problem  after processing the data from social media to estimate the spatio-temporal distribution of shortages for gasoline. Since one is traveling to find gasoline, the costs incurred in searching for the next location depend on the previous location visited.  The methods proposed in \cite{teller2019minimizing} are used for the search problem, and the overall model is demonstrated on a case study to obtain managerial insights.

In \cite{hellerstein2017max}, a variation of the STP is considered for $k$-out-of-$n$ functions. The problem is called max throughput and the goal is to maximize the throughput of a system for $k$-out-of-$n$ testing in a parallel setting where each test is performed by a separate “processor”. In this problem, in addition to the probabilities $p_i$, there is a rate limit $r_i$ associated with the processor that performs test $i$, indicating that the processor can only perform tests on $r_i$ items per unit time. The testing framework is also different in this paper. The testing continues until $k$ variables with value 0 are found or all variables are tested. A polynomial time exact algorithm is provided, generalizing previous results for parallel functions.

In \cite{elsayed2009port,boros2008optimization}, the inspection problem of incoming containers at a port by using sensors is studied. In this case, the sensors are not perfect, so the total cost includes misclassification costs as well as inspection costs. The imperfectness is modeled using a threshold, and the result of a test is assumed to be positive if the reading from the sensor exceeds the threshold. So, this is another parameter that can be tuned to find a good trade-off between the different cost categories. The underlying function is the parallel function.  In a follow-up study, in \cite{young2009multiobjective}, the inspection time is also taken into account, and a multiobjective version of the problem is studied.

In \cite{gabbert2013sequential}, a variation of the  STP is studied in a totally different application area. In particular, the goal is to find out the level of toxicity of a substance by applying costly tests that are not perfect. The common concept is that they try to construct a decision tree considering the costs, sensitivity, and specificity of the tests. A value of information-based approach is used to solve the problem.
In \cite{chi2010decision}, a medical application is considered. The problem is quite different than the STP in the sense that it is not the case that we have a given Boolean function. Instead, there are some example results of tests and diagnosis by doctors from previous cases. The problem is to develop a "good" testing policy in terms of cost, speed, and accuracy. Accuracy is measured with respect to the given instances. The problem is solved by an approach that uses machine learning techniques. The similarity with the STP is that they try to come up with a decision tree. Similarly, in \cite{park2011cost}, a learning technique referred to as cost-sensitive case-based reasoning is used for medical diagnosis. In \cite{ling2006test}, the authors develop cost-sensitive machine learning algorithms to model this learning and diagnosis process considering testing costs and misclassification costs. Various heuristic algorithms including sequential, single batch, and multi-batch  are tested on a case related to heart disease.  Other examples in medical testing include \cite{albu2017logical}, where they consider building a decision tree of tests for hepatitis B, and \cite{azar2013decision} where they consider breast cancer. Although many papers motivate the problems using examples from medical diagnosis, it turns out there have been no good examples of results published that bring together the domain knowledge from medicine and the algorithmic developments for the STP. This is a potential area for further research.

\cite{baron2024decentralized} considers the problem of sequencing stores to fulfill an online order consisting of different items. These types of problem have been widely studied in the literature, yet this is the first to set up a framework similar to the STP.  Each store may accept to ship none or some of the items in the order independent of other stores with certain probabilities that depend on the number of items left in the order. The cost structure in this problem is somewhat different in the following way.  If a store rejects all items on an order, no shipping cost is paid.   On the other hand, the retailer pays a fixed cost to a store if the store accepts to ship some of the items in the order. Once the whole order is completed, it is shipped to the final customer. The problem for the retailer is to find an optimal order of the stores to ask for the items, minimizing the expected cost of fulfilling orders. Essentially, the underlying function describes whether the order is completed or not. The stores are asked one by one if they can provide the items in the order.  Different cases have been considered such as when no splitting of orders is allowed or there are two items in the order, two-store case etc. For relatively simple cases, closed form solutions can be obtained. For more complex cases, different heuristic policies have been proposed and numerically evaluated.

In \cite{agnetis2009sequencing}, a scheduling problem is studied on parallel machines where a benefit is obtained when a job is successfully executed, and once a job has failed, the machine is blocked and no benefit is obtained for the jobs scheduled after the failed job. The goal is to find the allocation of jobs to machines and the sequence of jobs on each machine to maximize the total expected benefit. It is shown that the problem is NP-complete when there are 2 machines.

\cite{kowshik2011information,kowshik2013optimal}, a wireless sensor network is considered in which the transmission of each node can be heard by all other nodes. Optimal strategies are developed for computation and communication in the network in such wireless sensor networks.  Different types of networks are considered, and all of the nodes compute a Boolean function. Average-case and worst-case complexity of computing the Boolean function is for threshold, interval function when the network is out-tree and collated.

\section{Conclusion}\label{conclusion}
The most significant results for STPs have been approximation algorithms and hardness results for different classes of Boolean functions in the last 20 years. Still, there are numerous open problems in this respect, such as developing algorithms with better approximation bounds or considering different functions. In spite of all these results that utilize very different techniques, for some classes of functions, there have not been significant developments, read-once functions being an example. Another avenue of research is special cases of hard problems.  Some new extensions of the problem have also been studied, such as the case with precedence constraints, batch testing. Lately, SSCP and SHEP have attracted attention as new extensions. Further extensions are expected to emerge as a consequence of other application areas such as machine learning, medical diagnosis, etc.  Domain knowledge and the latest algorithmic developments in the area should be utilized together to obtain impactful results.   Practical heuristic algorithms have also been proposed in the literature for certain application areas. However, application papers that use real data are quite rare.  Another possible research direction is to utilize robust optimization techniques with respect to the cost and the probabilities. This would be a natural extension for all types of functions since these parameters are typically estimated before the actual realizations. One can also consider a more general cost structure where the cost of learning a variable depends on the previously learned variables, which would be relevant in applications where a cost is incurred before conducting the next test, such as checking network connectivity by deploying equipment or a team.


\bibliographystyle{elsarticle-num} 
\bibliography{review2024_revised2}





\end{document}